# MODEL REDUCTION METHODOLOGY FOR COMPUTATIONAL SIMULATIONS OF ENDOVASCULAR REPAIR.

**V.A. Acosta Santamaría**[1,4,5], **G. Daniel**[1,3,4,5], **D. Perrin**[1,4,5],
**J.N. Albertini**[1,2,5], **E. Rosset**[1,3] **S. Avril**[1,4,5]

[1] SaInBioSE, INSERM, U1059, F-42023 Saint Etienne, France
[2] Centre Hospitalier Universitaire de Saint-Etienne, F-42023 Saint Etienne, France
[3] Centre Hospitalier Régional Universitaire de Clermont-Ferrand, F-63000 Clermont-Ferrand, France
[4] Mines Saint-Etienne, F-42023 Saint Etienne, France
[5] Université de Lyon, F-42023 Saint Etienne, France

victor.acosta@emse.fr

## ABSTRACT
Endovascular aneurysm repair (EVAR) is a current alternative treatment for thoracic and abdominal aortic aneurysms, but is still sometimes compromised by possible complications such as device migration or endoleaks. In order to assist clinicians in preventing these complications, finite element analysis (FEA) is a promising tool. However, the strong material and geometrical nonlinearities added to the complex multiple contacts result in costly finite-element models. To reduce this computational cost, we establish here an alternative and systematic methodology to simplify the computational simulations of stent-grafts (SG) based on FEA. The model reduction methodology relies on equivalent shell models with appropriate geometrical and mechanical parameters. It simplifies significantly the contact interactions but still shows very good agreement with a complete reference finite-element model. Finally, the computational time for EVAR simulations is reduced of a factor 6 to 10. An application is shown for the deployment of a SG during thoracic endovascular repair, showing that the developed methodology is both effective and accurate to determine the final position of the deployed SG inside the aneurysm.

## 1. INTRODUCTION
The aorta can present pathological wall dilatation, called aneurysm. For treating abdominal aortic aneurysms (AAA), descending thoracic aneurysms (DTA), thoracic transections or patient with acute type B dissections, the EVAR treatment is an alternative to open surgery repair (OSR) [1][2][3][4][5][6][7]. EVAR is a minimally invasive technique, and consists in the insertion and deployment of one or more SG to prevent further enlargement or the aneurysm rupture. The intervention consists in the insertion of a SG inside the aneurysm via femoral artery [1][2][5][8]. There are currently over 100 different types of stent grafts in the market and laboratories in the world [9]. Treatment with EVAR is sometimes compromised by mechanical issues like an inadequate location of the SG, device compression, graft kinking, fatigue failings or aortic perforation [1][5]. The post-operative complications could be endoleaks, device migration, stenosis or thrombosis [1][8][10].

These last years, the Finite Element Method (FEM) has been extensively used to study various devices for the treatment of cardiovascular diseases [1][9][11][12]. FEA appears to be a quick and cheap method to evaluate: (i) the mechanical response to angioplasty and stent placement inside arterial walls, (ii) the risk of aorta rupture, (iii) the mechanical effects of stent deployment, (iv) the interactions between balloon–stent and stent–plaque–blood vessel systems, (v) the effective yielding limit of the structure [1][4][9][11][12][13][14]. The SG interaction with the intraluminal blood has been investigated using Computational Fluid Dynamics (CFD) and Fluid Structure Interaction (FSI) modeling [1][7].

The mechanical behavior of the stent includes geometric and material non-linearities and complex multiple contact behaviors, which are challenging for numerical simulation [9][15]. The presence of the textile onto which stents are sutured is a key aspect that drastically influences the device behavior and requires specific modeling. In previous studies, simplified SG models have been defined [10][16][17][18]. Moreover, these models did not take into account the mechanical interactions between stents and graft. The feasibility of FEA to simulate deployment of marketed SG in curved aneurysm models was demonstrated in a previous study of our group [10][19]. The goal of the present study is to define an alternative model reduction methodology for Thoracic Endovascular Aortic Repair (TEVAR) computational simulations. To reduce the computational time, a simplified but representative model of a real deployed SG is proposed. The simplified model has to reproduce the same mechanical behavior of the real SG model. The optimal parameters of the equivalent simplified model are defined by the *Robust Design Method*. Finally, the developed methodology is successfully evaluated for the simulation of SG deployed in a real TEVAR intervention.

## 2. MATERIALS AND METHODS
### 2.1 Methods
#### 2.1.1 Equivalent orthotropic elastic material parameters

An important aspect for the simplified approach was to determine the equivalent orthotropic elastic material parameters that represent the same average mechanical behavior as the selected real SG model. A linear elastic orthotropic behavior was assumed. In order to obtain the stiffness parameters, a general expression for the internal strain energy of a linear elastic structure was considered. If the stresses and strains are written as vectors $\{\sigma\}^T$ and $\{\varepsilon\}^T$, the expression can be written compactly as:

$$U(t) = \frac{1}{2}\int_0^t \int_V \{\sigma(t)\}^T \{\dot{\varepsilon}(t)\} dV dt \qquad (1)$$

Where, $U$ is the total strain energy and the entire volume is $V$ (associated to SG).

To evaluate the equivalent stiffness of the structure three different loading types were applied: radial compression, axial tension and torsion. For each loading type, the total strain enrgy $U_{radial}$, $U_{tension}$ and $U_{torsion}$ was calculated. These energies were used to scale the stiffness parameters of the simplified model (SM), using the following energy linear approximation:

$$U = \frac{1}{2} E \varepsilon^2 Vol \qquad (2)$$

Where, $E$ is the corresponding modulus for the defined type of loading. From the radial compression loading, it was possible to scale the $E_1$ orthotropic modulus, the $E_2$ modulus was scaled from the axial tension loading type and the $G_{12}$ modulus from the torsion condition. The volume ($Vol$) was defined with the SM geometry and considering an arbitrary equivalent thickness ($t$).

### 2.1.2 Equivalent membrane behavior and bending stiffness calibration

Equivalent stiffness parameters being deduced for an arbitrary thickness $t$, provided a suitable membrane behavior of the shell. However, the bending stiffness of the shell still had to be calibrated. This can be understood by introducing the theory of classical laminates, where the complete set of constitutive equations can be summarized as a single matrix equation:

$$\begin{Bmatrix} N \\ M \end{Bmatrix} = \begin{bmatrix} A & B \\ B & D \end{bmatrix} \begin{Bmatrix} \varepsilon^0 \\ K \end{Bmatrix} \tag{3}$$

Where $N$ and $M$ represents the tensions and bending moments applied to the plate. The $A/B/B/D$ matrices define the laminate stiffness matrix. $A$ matrix gives the influence of an extensional mid-plane strain ($\varepsilon^0$) on the in-plane tension $N$. $B$ is a coupling matrix between the membrane and the bending behavior. $D$ is the bending stiffness matrix. $K$ is the vector of second derivatives of the deflection (curvatures). They all depend on $t$ and the stiffness parameters ($E_1$, $E_2$ and $G_{12}$). Traditionally, $D = A(t^3/12)$ and only $t$ has to be calibrated to set the correct bending behavior when the membrane behavior has already been identified.

Due to the nonlinear geometric effects of SG's, it was necessary to define an optimal set of parameters that would provide an equivalent membrane and bending behaviors. With the initial parameter values ($E_1$, $E_2$ and $G_{12}$) and considering an arbitrary initial $t$ and a Poisson ratio ($v_{12}$), we explored their neighborhood through the design of experiments approach (DoE). The objective was to find the correction factors (CF) for each parameter that would permit to achieve the best agreement between SG and SM (considering a compression loading case).

The *Robust Design Method* provide an efficient approach to determine the optimal configuration of the parameters [20][21]. In order to reduce the number of observations, the *Taguchi Methodology* is usually applied using fractional factorial designs [22][23][24]. The necessary combinations are expressed as $n_{(levels)}^{k(factors)}$. It was necessary to define $n \geq 3$, to consider the possible non-linear effects of the trends. For the definition of levels, it was important to consider that the total work done by all strain components remained positive and the value $1 - (E_2/E_1)v_{12}^2$ remained greater than zero [25]. The target was the agreement between the applied displacement ($u_{compression}$) and the reaction force (RF) for SG and SM models. Statistical data processing was applied on the RF results, to determine: the effects of the factors, the interactions between factors and the influence percentage of each factor with respect to the target variable [23][22].

### 2.1.3 Material parameter calibration

A predictive function was proposed to evaluate and select the appropriate factors, applying the *Multiple Linear Regression Method*. The function involved: the evaluated factors, the non-linear effects and the synergic interactions that were reported. A normality test was applied on the reported data to find outliers. Finally, to calibrate and determine CF of the orthotropic mechanical parameters for SM, the *Response Optimizer Regression* was applied.

### 2.1.4 Statistical analysis

A parametric inference was assumed for the DoE results, and a normal distribution was verified by the *Anderson-Darling* test (p-value ≥0.05). The analysis of variance (ANOVA) was implemented to evaluate the influence percentage of the factors ($E_1$, $E_2$, $G_{12}$, $v_{12}$ and $t$). The *Multiple Linear Regression Method* was applied for the definition of the predictive function, using the *Stepwise Regression* (confidence interval level of 95%). The predictive function was evaluated with the standard distance that data values fell from the regression line (*S*), the proportion of the variation in the response data explained by the model ($R^2$) and how well the model predicted data ($R^{2pred}$). Finally, the *Response Optimizer Regression* was used to calibrate the orthotropic mechanical parameters for SM. Those statistical methods were performed using Minitab®.

## 2.2 Application
### 2.2.1 Clinical case – real stent-graft computational model

A clinical case was selected from a group of patients treated for type B thoracic dissections (University Hospitals of Saint-Étienne and Clermont-Ferrand - France). The main feature of this case was a highly compressed and narrow true lumen due to a largely thrombosed false lumen (80 mm of maximum diameter). Surgical intervention consisted in the deployment of a thoracic SG in the true lumen with a coverage length of 217 mm from the left subclavian artery. The implemented SG (Zenith – Alpha Thoracic®), is used for TEVAR treatment and to treat Thoracic Aortic Aneurysms (TAA). The pre-operative and post-operative CT-scans were obtained (see figure 1).

To extract the true lumen model from the pre-operative CT-scans, the CT-scan data was imported using the Crimson software [26], see figure 1c. The final geometry was exported as an IGES file, and imported in Abaqus to define a structural mesh. The true lumen geometry was considered as a rigid body and meshed with 3456 linear 4-node shell elements (S4R), sized 2.5 mm (see figure 1d). Additionally, the entire SG model was defined by assembling nine real stent-graft segments (SG-S), see figure 2d).

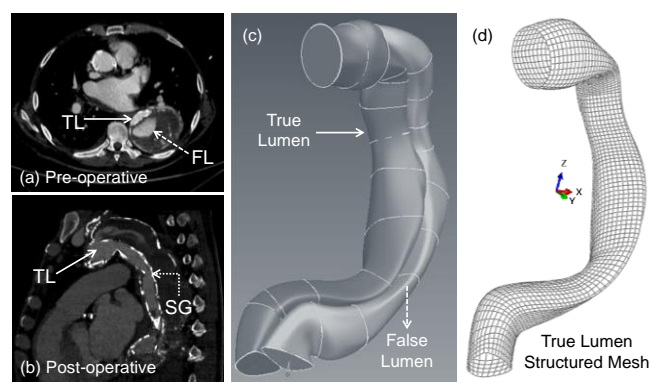

Figure 1. Image processing method applied to obtain the aortic model. (a) Pre-operative CT-scans. (b) Post-operative CT-scans. (c) Segmented models of the aortic lumens (true and false). (d) Structured mesh defined for the true aortic lumen.

3D digitized geometry of one SG-S and the mechanical parameters of Polyester fabric and Nitinol were provided by the manufacturer, see table 1 and figure 2b. The reference finite-element model for the SG was extensively validated against experimental data during previous studies conducted in our group [10][27][28][29]. For SG-S, the graft was modeled as an orthotropic linear elastic material and meshed with linear 4-node shell elements (S4R), see table 1. Additionally, and considering that the superelastic behavior of Nitinol remained in its austenitic phase during SG deployment, and according to measured austenitic modulus, the Nitinol stent was modeled as a linear elastic material [19]. The stent was meshed with linear beam elements (B31), see table 1. The SG-S entire volume ($V$) is equal to 125.85 mm$^3$ (graft$_{volume}$ + stent$_{volume}$), see table 1. As the stent and the graft are actually sutured together, specific tie constraints were prescribed to represent the suture points preventing the stent and graft from sliding or separating during simulations, see figure 2a - 2b.

### 2.2.2 Simplified stent-graft computational model

The proposed SM represent the interaction area between the graft and the stent in the SG-S, see figure 2c – 2e. The SM was defined as a shell geometry. The diameter of the stent and the thickness of the graft, in the SG-S model, were assumed as the initial equivalent thickness ($t$ = 0.35 mm) and the volume ($Vol$) was 116.79 mm$^3$ (see table 1). The SM was meshed with the same mesh properties as the ones used for the graft in the SG-S model, see table 1. The assigned mechanical properties corresponded to the orthotropic elastic parameters defined in the present methodology ($E_1$, $E_2$, and $G_{12}$). In addition, an arbitrary initial $v_{12}$ of 0.3 was defined. Finally, for the clinical case, the entire SM consisted in assembling nine simplified models (SM's) and eight shell sections that worked as graft connectors, see figure 2e.

Table 1. Manufacturer mechanical properties for the real stent-graft segment (SG-S).

| SG-S | Material | $E_1$ [MPa] | $E_2$ [MPa] | $G_{12}$ [MPa] | $v_{12}$ | $t$ [mm] | Volume [mm$^3$] | Element Type | Element Size [mm] | Number of Elements |
|---|---|---|---|---|---|---|---|---|---|---|
| Graft | Polyester | 1125 | 5000 | 18 | 0.2 | $t$ = 0.02 | 116.79 | Shell | 0.2 | 9996 |
| Stent | Nitinol | 60000 | - | - | 0.33 | Ø = 0.33 | 9.06 | Beam | 0.23 - 0.56 | 432 |

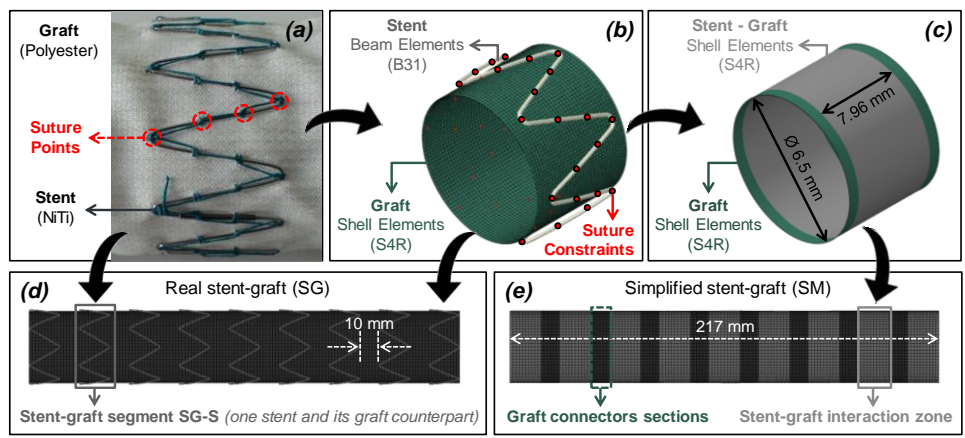

Figure 2. Computational models. (a) Real stent-graft segment (Zenith – Alpha Thoracic®). (b) Real stent-graft segment model (SG-S). (c) Simplified stent-graft model (SM). (d) Entire stent-graft clinical model (SG). (e) Entire simplified stent-graft clinical model (SM).

## 2.3 Methodology.

Once a commercial SG was selected, a real SG-S was considered (one stent ring and its graft counterpart), see figure 2a – 2d. The SG-S was implemented only for the proof of concept and for the validation analysis. For this structure, the average displacements and total internal strain energies were calculated to define the initial equivalent orthotropic elastic material parameters for SM ($E_1$, $E_2$, and $G_{12}$), and considering three loading conditions (radial, tension and torsion). The computational models were solved with a standard solver.

To adjust the equivalent membrane behavior and the bending stiffness of SM, the *Robust Design Method* was applied using the *Taguchi Methodology*. In the present study the orthogonal array L'$_{16}$ was implemented to evaluate five factors ($E_1$, $E_2$, $G_{12}$, $v_{12}$ and $t$), and with four different levels of analysis ($4^5$). To define the final set of levels, a preliminary sensitivity study was developed using an additional L'$_{16}$ arrangement (see table 2). The RF, reported by SM, was the target for the DoE. The applied loading case was a compression test (8 mm). To solve the L'$_{16}$ array, sixteen different computational simulations were required (standard solver). Finally, and for the material parameters calibration, a predictive function was determined in order to define CF for each parameter. Three additional computational cases were defined to solve the *Multiple Linear Regression* for an orthogonal array L'$_{16}$. The selected levels for these cases were the same order for all factors (model 17 = level 2, model 18 = level 3 and model 19 = level 4), see table 2.

For the clinical case, and regarding the real SG, the final equivalent parameters were applied in a corresponding SM in order to compare the deployed behavior. The methodology used to perform the reference simulation was directly derived from the numerical process detailed and validated by Perrin et, al. 2015 and 2016 [19][30]. First, a morphing algorithm computed nodal displacements to morph the original mesh of the arterial lumen to a cylinder geometry, while maintaining the same mesh topology (see figure 3a). The crimped SG was then introduced inside the computed cylindrical mesh, its longitudinal position being adjusted to obtain the targeted proximal SG positioning after completing the deployment simulation (see figure 3b). The next step consisted in a finite-element simulation (explicit solver), where the cylindrical mesh was morphed back to the pre-operative arterial lumen geometry (using the displacements computed in the first step), see figure 3c. While performing this simulation, contact constraints were activated between the cylindrical mesh and the SG model, leading to the deployment of the SG inside the pre-operative geometry. The same technique was applied for the SM model. The applied friction coefficient was the same for both models (0.15).

For the computational models and to establish the appropriate element size, a mesh independence study was conducted in order to guarantee that the results were grid independent. To the computational models solved with the explicit solver, the time steps and mass scaling were adjusted to obtain fast results, while ensuring that the ratio of kinematic and internal energy was not higher than 10% and that no spurious dynamic effects were observed in the results. The computational models were solved with Abaqus® (Simulia, Dassault Systems, Providence, RI, USA).

Table 2. Definition of factors, operational ranges and levels used in the orthogonal arrangements L'$_{16}$ ($4^5$).

| | Orthogonal Array L'$_{16}$ - ($4^5$) five factors and four levels | | | | | | | |
|---|---|---|---|---|---|---|---|---|
| Factors ($k$) | Levels ($n$)- Preliminary Sensitivity Study | | | | Levels ($n$)- Final Study | | | |
| | 1 | 2 | 3 | 4 | 1 | 2 | 3 | 4 |
| Modulus $E_1$ [MPa] | 1100 | 1467 | 1833 | 2200 | 1280 | 1346.67 | 1413.33 | 1480 |
| Modulus $E_2$ [MPa] | 11100 | 11833 | 12567 | 13300 | 12000 | 12166.67 | 12333.33 | 12500 |
| Modulus $G_{12}$ [MPa] | 6 | 2004 | 4002 | 6000 | 50 | 200 | 350 | 500 |
| Poisson's ratio $v_{12}$ | 0.28 | 0.29 | 0.3 | 0.31 | 0.28 | 0.29 | 0.3 | 0.31 |
| Thickness $t$ [mm] | 0.32 | 0.34 | 0.36 | 0.38 | 0.32 | 0.34 | 0.36 | 0.38 |

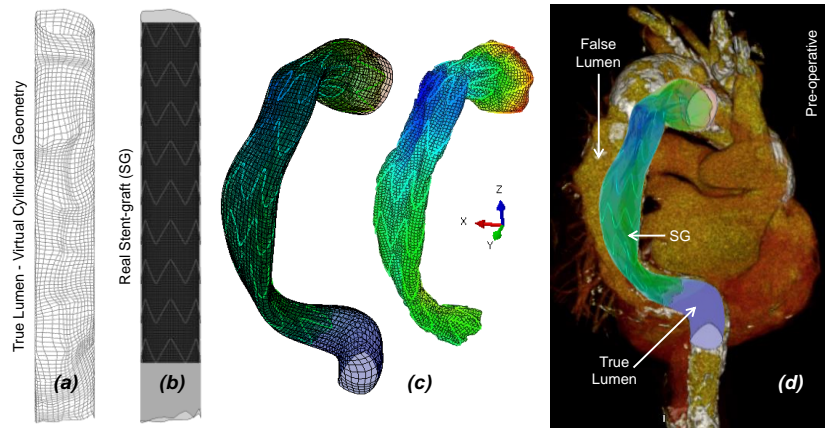

Figure 3. Mesh morphing methodology applied to define the mechanical deployed behavior and the final position of the real stent-graft model (SG). (a) Definition of the virtual cylindrical geometry for the true lumen. (b) The crimped SG was positioned inside the cylindrical mesh and the longitudinal position was adjusted. (c - d) A finite-element model defined the SG spatial configuration with regard to the pre-operative CT-scans.

## 2.4 Validation.

For the validation, three different aspects were considered. The first one was the validation of the equivalent membrane behavior and the bending stiffness of SM. This comparison was performed using a compression loading case. The RF's reported values were evaluated between SG-S and SM models and the predictive function. Additionally, the displacement behaviors of the computational models were compared (SG-S and SM). The second feature was the estimation and comparison of the computational time required by the models (SG-S and SM). Moreover, the computational time was estimated for the considered clinical case (real and simplified models).

Besides reducing the computational time, the goal of the reduction methodology was to predict the deployed position of the SG inside the aneurysm. The final position for the corresponding SM was compared with the SG computational model. However, and regarding to the post-operative CT-scans, the final stent locations were verified in the SG computational model. From the post-operative CT-scans, the stents were segmented and aligned with the computational results using ImageJ software [19]. The position of each point corresponding to each stent was derived using a script in Matlab®.

## 3. Results

Because the stent and the graft are sutured together, it was necessary to consider the pre-stress induced by the suturing process in the SG-S finite-element model. A preliminary FEA was performed to define the pre-contact between the stent and the graft. Simulation started with the oversized stent. To define the pre-contact, a radial displacement was assigned to the stent ($u_{pre-contact}$ = 0.497 mm), see figure 4a. The maximum pre-stress reported was 697.9 MPa (Von Mises).

The SG-S was used to define the displacements ($u$) under different loading conditions. For the radial compression loading, the final displacement ($u_r$) was equal to 0.34 mm. To this condition the average displacement obtained in the pre-contact state was cancelled ($u_r = u_{radial} - u_{pre-contact}$). For the axial tension and torsion loadings, the displacements were $u_{tension}$= 0.07 and $u_{torsion}$ = 0.12 mm (see figure 4). About the total strain energies, the highest value was established for the radial displacement ($U_{radial}$= 208.91 N/mm²). Additionally, the corresponding total strain energies $U_{tension}$ and $U_{torsion}$ were 57.89 and 69.22 N/mm², respectively. For SM, the initial equivalent orthotropic elastic material parameters were calculated, see table 3.

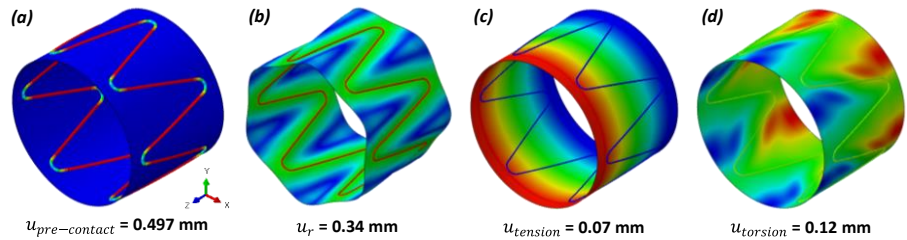

Figure 4. Average displacements ($u$) reported for four different loading conditions: pre-contact, radial, tension and torsion.

The *Taguchi Methodology* was used to explore the optimal parameters (assuming a compression loading). About the estimation of effects of the factors and with respect to RF, $G_{12}$ was the only factor that showed a linear trend ($R^2$ = 0.98). When the levels of the factors increased, the RF increased ($E_2$, $G_{12}$, $v_{12}$ and $t$). An opposite trend was found for the factor $E_1$, the RF decreased when the levels increased. The relationship between the different factors showed a synergic interaction, all the factors were considered as dependent variables. Finally, the factor influence showed that $v_{12}$ and $G_{12}$ were the most relevant factors on the RF behavior (25.7% and 25.6% of influence). The influence for the factors $E_1$, $E_2$, and $t$ were 24.2%, 10.5% and 14%, respectively.

For parameter calibration, a *Multiple Linear Regression* was implemented. In the present study, the RF value reported for the fourth observation in the arrangement L'₁₆ was considered as outlier (2.44 N/node). If this value was omitted, the normality test established a normal distribution (p-value = 0.185). The *Stepwise Regression* defined the next predictive function for RF ($S$ = 0.07, $R^2$ = 92.82% and $R^{2-pred}$ = 81.03%), see figure 5.

$$RF = -4.42 + (E_1 * -1.67E^{-3}) + (E_2 * 2.54E^{-4}) + (G_{12} * -3.25E^{-3}) + (v_{12} * 13.4) + (G_{12}t * 0.012) \quad (4)$$

The *Response Optimizer Regression* was applied to determine the correction factors of the orthotropic mechanical parameters. RF was set to the data reported for the SG-S model, under the compression loading (0.63 N/node). The $v_{12}$ parameter presented the highest influence and consequently this value was set constant ($v_{12}$ = 0.3). The optimization determined the final mechanical properties for SM, see table 3. Finally, and using the interaction $G_{12}t$ (23.69 MPa mm, for the present study) the virtual thickness was adjusted ($t$ = 0.474 mm), see table 3.

Table 3. Orthotropic elastic parameters and correction factors defined for the stent-graft simplified model (SM).

| Simplified Model (SM) | $E_1$ [MPa] | $E_2$ [MPa] | $G_{12}$ [MPa] | $v_{12}$ | $t$ [mm] |
|---|---|---|---|---|---|
| Initial parameters | 1342.78 | 13164.85 | 5356.17 | 0.3 | 0.35 |
| Final parameters | 1280 | 12008.32 | 50 | 0.3 | 0.474 |
| Correction Factors | 1.05 | 1.10 | 107.12 | 1 | 0.74 |

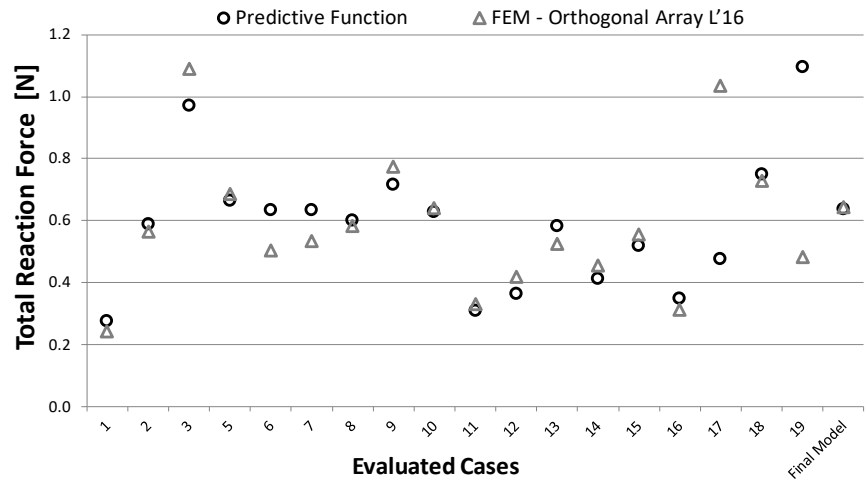

Figure 5. Data reported by the orthogonal array L'$_{16}$ under a compression loading, for the simplified finite-element model (SM) and the predictive function.

## 3.1 Validation

The new parameters were implemented in a final computational model to verify the membrane and bending stiffness behaviors of SM. The RF reported with the predictive function was 0.638 N/node and 0.644 N/node for SM, the relative error was 0.93% (see figure 5). For the SG-S, the RF reported value was 0.632 N/node. About the SG-S and SM models, the relative error was 1.95%. Finally, the average displacement for the computational models presented a relative error of 2.4% ($u_{\text{SG-S}}$ = 2.50 mm and $u_{\text{SM}}$ = 2.44 mm).

For the compression loading case, the computational time was reduced by 98% with regard to SG-S and using 8 calculation nodes in a high-performance computing cluster (see table 4). Another comparison was considered with a desktop personal computer. Similarly, the computational time was reduced by 98% (see table 4). Additionally, the real chronic type B dissection was used to evaluate the model reduction methodology defined in the present work. With respect to SG, the time was reduced by 86% with SM (see table 4). To SM, when the size of the elements decreases, the reduction of the computational time increases. For the clinical case application, another comparison was defined by decreasing the element size of the simplified model (from 1.58 to 0.9 mm). In the real model, the same parameter was modified for the graft mesh. The time was reduced by 62% (see table 4).

About the final deployed position, the comparison was performed between the post-operative CT-scans, SG and SM models (see figure 6). The SG mechanical behavior of the real model corresponded to previous results obtained in our group [10][28]. For the post-operative CT-scans, with respect to the pre-operative condition, the distal part of the true lumen showed the maximum displacement differences, see figure 6a. This difference was defined by the location of the SG deployed inside the aneurysm (post-operative CT-scans), see image 6a - 6b. For SM the maximum displacements reported a difference of 0.5%, concerning SG (see image 6c - 6d). Even if the displacements and the deployed behavior were similar for both models, the simplified segments showed a small sliding effect that was propagated from the proximal to the distal zone. This sliding generated an offset with respect to the real stents (see image 6e). The maximum offset was located in the distal part of the SG model ($X \pm 12.81$, $Y \pm 5.63$ and $Z \pm 7.43$ mm). Furthermore, the offset started on the proximal part ($X \pm 1.62$, $Y \pm 4.70$ and $Z \pm 1.50$ mm).

Table 4. Computational time reported by the real stent-graft model (SG) and the simplified stent-graft model (SM), for a compression test and the proposed clinical case (aortic dissection type B).

| Computer Specifications | Computational Time (min) | | | | | |
|---|---|---|---|---|---|---|
| | Compression Loading Case | | Clinical Case Mesh 1 | | Clinical Case Mesh 2 | |
| | SG-S | SM | SG | SM | SG | SM |
| Cluster (6 cores - 2.66 GHz - 24 GB RAM) | 52.20 | 1.01 | 27.99 | 3.91 | 49.03 | 18.41 |
| Desktop PC (4 cores - 3.5 GHz - 16 GB RAM | 132 | 2.04 | - | - | - | - |

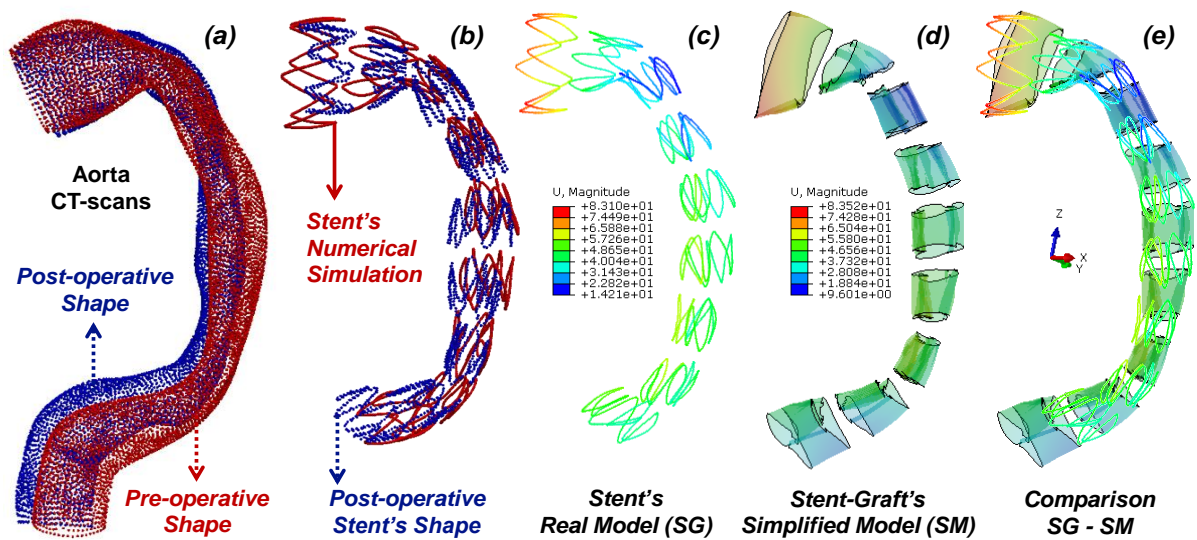

Figure 6. Final predicted position of the deployed stent-graft inside the TAA. (a) Pre and post-operative CT-scans of the true lumen. (b) Comparison between the segmented stents (post-operative CT-scans) and the SG model.

(c) Displacements behavior of the stents (SG model). (d) Displacements behavior of the simplified sections (SM). (e) Comparison between the real and simplified models.

## 4. Discussion

Finite element analyses are very promising to assist surgeons for predicting the position of deployed SG's during TEVAR and EVAR interventions [19][30]. However, the computational cost of these FEA's remained a major issue. In the current work, an efficient model reduction methodology for endovascular repair computational simulations was proposed. As well as in the other numerical methodologies that can be applied to reduce the computational cost, the SG mechanical properties, interactions and geometric aspects were also considered [16][17][18]. These parameters changes depend on the SG's available in the market. In this context, these methodologies can be considered as deterministic methods.

D. Chen et all., (2013) defined a fast virtual stent-graft deployment algorithm based on contact mechanics, spring analogy and deformable meshes. For the authors, the SG was assumed as an integrative body. The wire thickness and stent cell shape were considered to define the elastic property of the stent mesh. Moreover, the mechanical behavior was assumed homogeneous. The interaction between the expanding SG and the vessel wall was evaluated (aortic dissection type Stanford B). The elastic properties of the aortic wall were involved. The required computational time was approximately 4 minutes. However, the initial length of the stent-graft was assigned the same length as that obtained in the post-operative image and the tangential displacement was not involved [17].

K. Spranger et al (2015) defined a novel fast virtual stenting method, based on a spring–mass model. The authors implemented a finite-element model to estimate the residual forces. The outcomes were used as a reference for optimizing the stiffness parameters of the numerical fast method (FM). A genetic algorithm was applied to calibrate and define the final parameters. The stiffness parameters assumed a linear elastic behavior. The stent and graft meshes were merged to define an anisotropic grid (using triangulation constraints). For the clinical case exposed, the computational time was reduced by 99.88% (FM = 19.44 s) [18]. However, the mechanical interaction between the stent and the graft was not assumed and the clinical case was not validated with a post-operative condition.

In the present work the reduction methodology consider: (i) the SG geometric configuration, (ii) the mechanical interactions between the stent and the graft (e.g. suture and pre-stress conditions), (iii) the SG mechanical properties, (iv) the contact behavior between SG and the aortic wall, (v) the mechanical behavior of the deployed SG. Finally, RF was used as a target variable to define and calibrate the equivalent orthotropic elastic material parameters for SM ($E_1$, $E_2$, $G_{12}$, $v_{12}$ and $t$). Once the estimation of the material parameters was established, it was possible to predict and/or confirm tendencies with the predictive function.

An additional goal of the reduction methodology was to predict precisely the final position of the deployed SG. Considering the established differences, it is relevant to expose the limitations of the present study. The mechanical properties of the aorta were not involved. To prevent the sliding effect, the friction coefficient had to be calibrated in the contact definition (between the true lumen and SM). For the implemented clinical case, the false lumen was not considered in the computational models. However, the false lumen may introduce a significant stiffness influence and/or represent an important geometric boundary condition for the SG deployment.

Further clinical cases should be considered to verify the predictions of additional simplified models. Moreover, a complex tortuosity condition can be considered with the angulations of iliac arteries in AAA's cases [27]. Other studies perform computational analysis on the patient-specific vascular models, combining accurate medical images and advanced CFD and FSI models. The studies addressed the assessment of post-operative hemodynamic conditions, the SG interaction with the intraluminal blood and stress reduction on the aneurysm wall by SG placement [1][5][7]. In this way, the simplified models proposed by the present methodology could be useful to simplify these computational simulations. Additionally, the results could be compared with the fast virtual stenting method (FVS) that combined simplex deformable models, hemodynamic conditions and CFD analysis [16].

## 5. Conclusions
A methodology was developed, validated and successfully applied to perform fast FEA simulations of TEVAR interventions. To reduce the computational time, the model reduction methodology determined equivalent model parameters of a simplified shell model using a *Robust Design Method*. Eventually, the computational time was reduced by 98% and 86% respectively compared to reference traditional FEA. Future work will focus on extending the computational predictions to long-term effects involving arterial remodeling occurring after TEVAR interventions.

## 6. Acknowledgments
The authors would like to acknowledge the French National Research Agency (ANR) for funding the Endosim project (grant ANR-13-TECS-0012). The authors are also grateful to the European Research Council for grant ERC-2014-CoG BIOLOCHANICS.

## 7. Conflicts of interest
None